\documentclass[twocolumn,showpacs,amsfonts,aps,prc,nofootinbib,floatfix]
{revtex4}

\usepackage{amsmath}
\usepackage{bm}
\usepackage{graphicx}

\usepackage{epsfig}
\newcommand{\beq}{\begin{equation}}
\newcommand{\eeq}{\end{equation}}
\newcommand{\bea}{\vspace{0.25cm}\begin{eqnarray}}
\newcommand{\eea}{\end{eqnarray}}

\newcommand{\ro}{\mbox{{\boldmath
$\rho$}}}

\newcommand{\qb}{\mbox{{\bf
q}}}
\newcommand{\kb}{\mbox{{\bf
k}}}

\def\lsim{\mathrel{\rlap{\lower4pt\hbox{\hskip1pt$\sim$}}
    \raise1pt\hbox{$<$}}}         
\def\gsim{\mathrel{\rlap{\lower4pt\hbox{\hskip1pt$\sim$}}
    \raise1pt\hbox{$>$}}}         

\begin{document}


\title{
On the use of running 
$\alpha_s$
in calculations of radiative energy loss
of fast partons in a quark-gluon plasma
}
\date{\today}

\author{B.G.~Zakharov}

\affiliation{
Landau Institute for Theoretical Physics, Russian Academy of Sciences,
Chernogolovka, Moscow region, 142432 Russia}

\begin{abstract}
The incorporation of running 
$\alpha_s$ for the gluon emission vertex in calculations of radiative
parton energy loss in a quark-gluon plasma is discussed. It is argued
that the virtuality scale for running 
$\alpha_s$ for induced gluon emission is determined by the square of the
transverse momentum of the emitted gluon rather than by the square of the invariant mass of the final two-parton state
often used in the literature.

\end{abstract}
%

\maketitle


An important feature of quantum chromodynamics (QCD) is the decrease of the
effective coupling constant 
$\alpha_s=g^2/4\pi$
with increase of the particle virtualities. 
This property plays a significant role in the
dynamics of parton showers in jet production in
hard processes. As is known, 
in the leading logarithmic
approximation
parton cascade allows a
probabilistic description  in terms of successive 
$a\to bc$ decays \cite{BasicsQCD}
with the distribution in Feynman's variable $x=E_b/E_a$
and the transverse momentum $k_\perp$ of parton $b$
given by the expression
\beq
{dw}=\frac{dk_\perp^2}{k_\perp^2}\frac{\alpha_s(k_\perp^2)}
{4\pi}P_{a}^{bc}(x)dx\,,
\label{eq:10}
\eeq
where $P_{a}^{bc}(x)$ is the splitting function for the 
$a\to bc$
transition in the DGLAP equation.
The use of Eq. (\ref{eq:10}) in Monte Carlo generators, e.g.,
PYTHIA [2] (together with the condition of angular
ordering for a soft region $x\ll 1$ \cite{angle_od1,angle_od2})) 
makes it possible to describe a huge amount of data on the jet physics.
The emission of gluons is the dominant process in 
evolution of jets in the soft region. It is important that
the virtuality in the argument of running $\alpha_s$ is 
determined only by the
transverse momentum of the emitted gluon, and is
independent of the longitudinal variable $x$ \cite{DDT,BasicsQCD}. 
For this reason, the emission of soft gluons at 
$x\ll 1$ is
independent of the energy of parton  $a$; i.e., the situation
in QCD is similar to the emission of soft photons in
the Low theorem in QED \cite{Low}. 
This property would be violated, e.g., when the
square of the invariant mass of the $bc$ system, $M^2_{bc}=k_\perp^2/x(1-x)$, 
is used for the characteristic virtuality in 
running $\alpha_s$. The expression 
$\alpha_s(k_\perp^2)$ in Eq. (\ref{eq:10}) for jets in vacuum can be
obtained within the diagrammatic technique in the
momentum representation after summation over the
masses of states to which the emitted gluon can transit \cite{DDT}.

The problem of the choice of the running charge
argument for the gluon emission becomes more
complicated in the case of a parton shower in the
medium. Such a situation occurs in the case of
jet production in collisions of heavy nuclei at
RHIC and LHC energies when a hot quark-gluon
plasma (QGP) is produced in the initial stage of an $AA$ collision at proper 
time $\tau_0\sim 0.5- 1$ fm. The development
of the parton cascade at times 
$\tau\sim \tau_0\div L_{QGP}$ (here, $L_{QGP}\sim (1- 2)R_A$
is the size of the QGP, where $R_A$ is the
nucleus radius) occurs in the QGP. The energy
losses of fast partons in the QGP result in jet quenching, 
which is, in particular, manifested in the strong
suppression in $AA$ collisions  of spectra of particles with high $p_T$
 (which is characterized by the nuclear
modification factor $R_{AA}$) observed at the RHIC and
LHC. The energy losses in the QGP for RHIC
and LHC conditions are due primarily to the induced gluon emission 
caused by multiple scattering of fast partons in the 
medium \cite{BDMPS1,LCPI1,LCPI2,GLV1,W1,AMY}.

In contrast to the shower cascade in vacuum, it is
reasonable to analyze radiation loss in the medium
in the coordinate representation in the noncovariant
perturbation theory because the description in terms of usual Feynman diagrams 
in the
momentum representation is 
impossible because 
of a huge number
of diagrams. In this case, all fast particles between collisions with constituents of the medium are described
by plane waves on the mass shell, and are not characterized by 
virtualities as in the Feynman diagram formalism. The virtuality of particles can be qualitatively
determined using the uncertainty relation $\Delta p\Delta L\sim 1$
from the size of the spatial region $\Delta L$ filled by a plane
wave between decays of particles or their rescatterings 
on particles of the medium (which are usually
modeled by the static Debye-screened color centers). At
present, jet quenching is usually analyzed using formulas for  the 
single-gluon spectrum under the assumption of independent 
gluon emission for multigluon
processes \cite{RAA_BDMS}. The single-gluon spectrum for massive
partons for arbitrary magnitude of the Landau-
Pomeranchuk-Migdal effect can be obtained within
the light cone path integral (LCPI) approach 
\cite{LCPI1,LCPI3,LCPI4}. The calculation of the single-gluon spectrum with
running $\alpha_s$ involves the problem
of the choice of the running coupling constant in the
decay vertex. The effect of the running coupling constant is particularly strong for LHC energies where
the jet energy range is much wider than that for RHIC
experiments.

The parton cascade in medium is not ordered in
the parton virtualities, as in the case of the Feynman diagrams
corresponding to cascading of partons in vacuum \cite{BasicsQCD}.
Consequently, the reasoning used in \cite{DDT} to determine the
virtuality scale for running $\alpha_s$
is inapplicable. For the induced gluon emission
in the case of not too strong the Landau-Pomeranchuk-Migdal effect, squares of the transverse
momenta of gluons at 
$L_f\lsim L_{QGP}$ (where $L_f\sim 2\omega/m_g^2$
is the coherence length for emission of a
gluon with the energy $\omega$ \cite{LCPI1}) are concentrated in a
region up to several units of $m_g^2$ , where $m_g$ is the gluon quasiparticle
mass in the QGP ($m_g\sim 400$
for RHIC and LHC conditions \cite{LH}). This estimate corresponds to the diffusion relation for the typical
transverse distance passed by a parton in the $\rho$ plane
on the longitudinal length $L$ \cite{LCPI1}:
\beq
\rho\sim \sqrt{L/\omega}\,.
\label{eq:20}
\eeq
At $L=L_f$, this gives $k_\perp\sim 1/\rho\sim m_g$. However, when
$L_f$ becomes much larger than the QGP size, Eq. (\ref{eq:20}) includes $L_{QGP}$ rather than $L_f$  \cite{Z_OA}. 
In this
regime, $k_\perp^2\sim m_g^2(L_f/L_{QGP})$, which can lead to significant virtualities in running $\alpha_s$
(particularly for a small QGP produced in $pp$ and $pA$
collisions \cite{Z_RPP,Gale1}).

Two methods are currently in use for incorporating the
running coupling in decay vertices when calculating radiative energy losses. The first method
involves $\alpha_s(k_\perp^2)$, whereas the second method includes
$\alpha_s(k_\perp^2/x(1-x))$, which corresponds to the squared 
invariant mass of the final two-parton state. The
first method was used in analyses \cite{RAA08,RAA11,RAA12,RAA13,Z_RPP}, 
which were
based on the LCPI formalism \cite{LCPI1,LCPI3,LCPI4}, and in 
\cite{Gale1,Gale2}, which were based on the generalization of the
AMY formalism \cite{AMY} to the case of a
finite-size QGP. The second method was used in the
well known CUJET model \cite{CUJET,CUJET3,CUJET3_1,CUJET3_2} based on the 
GLV formalism \cite{GLV1} for a thin medium.
The squared invariant mass of the two-parton
state (including the parton masses) for $Q^2$ in the
running charge within the GLV formalism was also used in 
\cite{MD0,MD1,MD2,MD3,MD4,MD5,MD6}. The use of
$\alpha_s(k_\perp^2/x(1-x))$ gives a steeper increase of the nuclear
modification factor with $p_T$, which is due to suppression of the 
induced gluon emission with
increasing energy of the initial parton (because of
a decrease in $x$ at a fixed energy of the gluon $\omega$ and,
correspondingly, an increase of the squared
invariant mass $k_\perp^2/x(1-x)$\,). This receipt gives better
agreement with the LHC data on $R_{AA}$, that show
a steep increase of $R_{AA}$ with the hadron transverse momentum.
\begin{figure} [t]
\vspace{.7cm}
\begin{center}
\epsfig{file=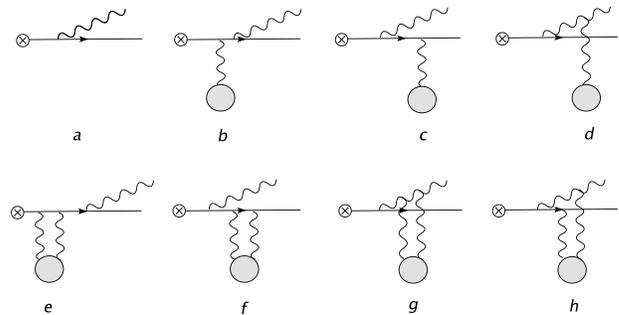,height=4.1cm,angle=0}
\end{center}
\caption[.]
{
Diagrams determining the $q\to gq$ transition
amplitude taking into account the single rescattering of
fast partons in a medium due to single and double $t$-channel gluon
exchanges with a constituent of the medium.}
\end{figure}
\begin{figure} [t]
\vspace{.7cm}
\begin{center}
\epsfig{file=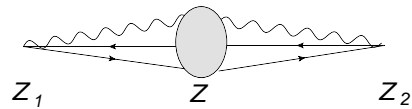,height=1.4cm,angle=0}
\end{center}
\caption[.]
{Diagrammatic representation for 
$d\sigma_{eff}^{BH}/dx$ in the
LCPI approach for scattering on a center at the point $z$.
Integration with respect to $z_{1,2}$ is performed over the
regions $0<z_1<z$ and $z_2>z$, respectively. The left and
right triangular parts stand for the Green's functions for
the Hamiltonian (\ref{eq:50}). The ellipsis stands for the three-body
cross section $\sigma_3$ given by Eq. (\ref{eq:80}).
}
\end{figure}
In this work, we  show that there are simple physical arguments 
against the use of $\alpha_s(k_\perp^2/x(1-x))$ in
calculations of the radiative energy losses of fast partons.

For definiteness, we consider the induced gluon emission for 
the $q\to gq$ process. It is
assumed that a fast quark is produced at $z = 0$ (the $z$
axis is chosen along the momentum of the initial
quark), and passes through a slab of the medium of
thickness $L$, which simulates the interaction in the
finite state for a jet produced in $AA$ collisions (where
$L\sim L_{QGP}$). The medium is described as a system of
color centers.
We consider first the case of a quite thin dilute
medium, when it is enough to account for only single rescattering 
of the fast partons on 
one of the color centers. The $q\to gq$ process with allowance for only
single interactions with constituents of the medium is
described by the diagrams shown in Fig.~1. In this
description, each fast particle is described by a plane
wave with a certain transverse momentum, which
changes sharply at rescattering on a color center with
exchange by one or two $t$-channel gluons \cite{Z_kin}. These
diagrams for arbitrary $x=\omega/E_q$ were calculated
in \cite{Z_kin}. The square of the sum of diagrams shown in
Fig.~1 after integration over transverse momenta gives
the gluon spectrum in the Feynman variable $x$ in the
form of the sum
\beq
\frac{dP}{dx}=\frac{dP_{vac}}{dx}+\frac{dP_{in}}{dx}\,.  
\label{eq:30}
\eeq
Here, the first term is the contribution from the
usual vacuum decay $q\to gq$ (diagram $a$ in Fig.~1), and
the second term (square of the sum of the diagrams $b$, $c$, $d$ in
Fig.~1 and the interference terms of the diagram $a$
and the sum of the diagrams $e$, $f$, $g$, $h$ in Fig.~1)
corresponds to the $q\to gq $ transition induced by the
interaction with a scattering center. The induced gluon spectrum 
corresponding to the diagrams in Fig.~1
can be represented in a compact form within the LCPI
approach \cite{LCPI1}. It corresponds to the 
leading in the medium density 
contribution to the total spectrum for
arbitrary number of rescatterings in the LCPI
method, and can be written in the form
\beq
\frac{d P_{in}}{d
x}=
\int\limits_{0}^{L}\! d z\,
n(z)
\frac{d
\sigma_{eff}^{BH}(x,z)}{dx}\,,
\label{eq:40}
\eeq
where $n(z)$ is the particle number density, and 
$d\sigma^{BH}_{eff}/dx$
is the effective Bethe-Heitler cross section including
the effect of finite size of the medium. In the LCPI
formalism, this cross section can be represented by the
single diagram shown Fig.~2, where ellipsis is the cross
section for the interaction of the color-singlet $q\bar{q}g$ system
with a  color center, $\sigma_{3}$,  and sets of three lines on the
right and left correspond to the Green's function for
the Hamiltonian
\beq
H=\frac{\qb^2+\epsilon^2}{2M}\,,
\label{eq:50}
\eeq
where $M=E x(1-x)$, and 
$\epsilon^{2}=m_{q}^2x^2+m_{g}^{2}(1-x)$ (generally, for the 
$a\to bc$ transition, 
$\epsilon^{2}=m_{b}^2(1-x)+m_c^2x-m_{a}^{2}x(1-x)$). The Hamiltonian 
(\ref{eq:50}) describes evolution in $z$ coordinate of the wavefunction of
the $gq$ pair in the $\rho$ plane (in this case, the antiquark
in the $q\bar{q}g$ system is located at the center of mass of the $gq$ pair
\cite{LCPI1}). Representing integrals with respect to $z_{1,2}$ in Fig.~2
in terms of the light cone wavefunctions in the $\rho$ representation, 
one can represent 
$\sigma_{eff}^{BH}(x,z)/dx$ in the form \cite{Moriond1998}
\beq
\frac{d\sigma_{eff}^{BH}}{dx}
=\frac{1}{2}\sum_{\{\lambda\}}\text{Re}\int d\ro \Psi^{*}_{\{\lambda\}}(\ro,x)
\sigma_3(\rho,x)
\Psi_{\{\lambda\}}^m(\ro,x,z)\,,
\label{eq:60}
\eeq
where $\{\lambda\}$ is the set of parton helicities, 
$\Psi_{\{\lambda\}}(\ro,x)$ is
the usual light cone wavefunction for the $q\to gq$ transition, 
and $\Psi_{\{\lambda\}}^m(\ro,x,z)$ is the light cone wavefunction
modified by the effect of finite size of the region of
the longitudinal coordinate $z_1$ for the
gluon emission vertices $0<z_1<z$.

The three-body cross section $\sigma_{3}$ can be
expressed in terms of the well known dipole cross section for a 
color-singlet $q\bar{q}$ pair \cite{NZ_sigma3}, which is given by
the formula
\beq
\sigma_{q\bar{q}}(\rho,z)=C_{T}C_{F}\int d\qb
\alpha_{s}^{2}(q^{2})
\frac{[1-\exp(i\qb\ro)]}{[q^{2}+\mu^{2}_{D}(z)]^{2}}\,,
\label{eq:70}
\eeq
where $C_{F,T}$ are the color quadratic Casimir operator
for the quark and thermal parton (quark or gluon), and
$\mu_{D}(z)$ is the local Debye mass
\footnote{Expression (\ref{eq:70}) is written in the form allowing 
for the use of running $\alpha_s$. The square of the momentum
through the gluon is a natural virtuality scale in the case of 
the $t$-channel gluon exchanges.}.
The three-body
cross section is represented in terms of the dipole cross
section (\ref{eq:70}) as
\bea
\sigma_{3}(\rho,x,z)=\frac{9}{8}
[\sigma_{q\bar{q}}(\rho,z)+
\sigma_{q\bar{q}}((1-x)\rho,z)]\nonumber\\
-\frac{1}{8}\sigma_{q\bar{q}}(x\rho,z)\,.
\label{eq:80}
\eea

To discuss the running coupling constant at the
$q\to gq$ vertex, it is convenient to represent Eq. (\ref{eq:60}) in
the momentum representation:
\bea
\frac{d\sigma_{eff}^{BH}}{dx}
=\frac{1}{2(2\pi)^4}\sum_{\{\lambda\}}\text{Re}\int d\kb_1 d
\kb_2\Psi^{*}_{\{\lambda\}}(\kb_2,x)\nonumber\\
\times\sigma_3(\qb,x)\Psi_{\{\lambda\}}^m(\kb_2,x,z)\,,
\label{eq:90}
\eea
where $\qb=\kb_1-\kb_2$. The wavefunctions for fixed $\alpha_s$
are given by the formulas
\beq
\Psi_{\{\lambda\}}
(\kb,x)=\sqrt{\alpha_s}\frac{(k_x-i\lambda_gk_y)
[2-x+2ix\lambda_q\lambda_g]}{\sqrt{2x}(\kb^2+\epsilon^2)}\,,
\label{eq:100}
\eeq
\beq
\Psi_{\{\lambda\}}^m(\kb,x,z)=F(\kb,x,z)\Psi_{\{\lambda\}}(\kb,x)\,,
\label{eq:110}
\eeq
where 
\beq
F(\kb,x,z)=1-\exp{\left[i\frac{(\kb^2+\epsilon^2)z}{2M}\right]}\,.
\label{eq:120}
\eeq
The running charge is introduced by changing in Eq. (\ref{eq:100}) the
fixed $\alpha_s$ to the running one. This exactly corresponds to the
use of running $\alpha_s$ for the
$q\to gq$ decay vertices in the diagrams shown in Fig.~1.
It is important that Eqs. (\ref{eq:60}) and (\ref{eq:90}) include not only
the square of diagrams with one-gluon exchanges in
Fig.~1 but also the interference between the vacuum diagram
$a$ in Fig.~1 and the diagrams with rescattering of fast partons on the color 
center because of two-gluon exchanges
(diagrams $e$, $f$, $g$, $h$ in Fig.~1). Only with this interference
terms can all $t$-channel exchanges be summed
in the form of the cross section $\sigma_3$, and the formula for
the effective Bethe-Heitler cross section can be
obtained in the simple factorized form (\ref{eq:60}). To keep the
form of Eqs. (\ref{eq:60}) and (\ref{eq:90}) at their generalization 
to the case of the running charge, running $\alpha_s$
coinciding with $\alpha_s$ in the vacuum diagram in
Fig.~1а, i.e., $\alpha_s(k_\perp^2)$ (if it is accepted that the vacuum
cascade includes this form of running $\alpha_s$ \cite{DDT}), 
should be used for diagrams with rescatterings in Fig.~1. The 
requirement for keeping the factorized form (\ref{eq:60}) is important
because vanishing of $\sigma_3(\rho,x,z)$ at $\rho\to 0$
(related to
the property of the color transparency of the dipole
cross section (\ref{eq:70}), i.e., vanishing of 
$\sigma_{q\bar{q}}(\rho,z)$ at
$\rho \to 0$) ensures the convergence of the $\rho$-integral in
Eq. (\ref{eq:60}). Form (\ref{eq:60}) implies the absence of the 
interaction with the medium for a point color-singlet $q\bar{q}g$
state. The induced transition physically appears
because $t$-channel gluons can distinguish, for a nonzero $\rho$, 
the two-body virtual state $qg$ and the
initial quark because of the color dipole moment of
the $qg$ pair. For this reason, at phenomenological
generalization of the formulas for the induced spectrum to the 
case of the running charge, it is reasonable
to require keeping of Eq. (\ref{eq:60}) having the property of color
transparency.

The above reasons in favor of the charge $\alpha_s(k_\perp^2)$
in the decay vertex are applicable when $L_f$ is
not too small as compared to the size of the medium,
and the contribution of interference with the vacuum
amplitude is significant. When $L_{QGP}$ is large, the contribution of color centers with $z\gg L_f$ dominates, and
the contribution to the induced spectrum from the
interference of the vacuum diagram with diagrams $e$, $f$, $g$, $h$ in
Fig.~1 should obviously be small. In this regime,
the induced gluon emission occurs in much the
same way as for the case of a quark incident on the scattering 
center from infinity, and the effective Bethe-Heitler cross section becomes close to the usual QCD
Bethe-Heitler cross section. However, the use of
$\alpha_s(k_\perp^2)$
for this regime also seems reasonable. Indeed,
the situation with $z\gg L_f$ is similar to the emission of
gluons in the dipole form 
\cite{NZZ_BFKL,NZ_HERA} of the BFKL equation \cite{BFKL1,BFKL2}. 
The analysis of radiative corrections to
the BFKL equation shows that the virtuality scale for
the running charge is determined only by the transverse momentum 
\cite{BFKL_NLO} (in agreement with the direct
calculations in the coordinate $\rho$-representation \cite{rBFKL}
where $k_\perp\sim 1/\rho$). In earlier works
\cite{BFKL_run,NZZ_BFKL,NZ_HERA},
 the running coupling constant was phenomenologically
incorporated in the BFKL equation also by using 
$\alpha_s(k_\perp^2)$ (with $k_\perp^2\sim 1/\rho^2$ 
for the dipole approach \cite{NZZ_BFKL}). With such
a running charge, a good description of the HERA
data on the proton structure function $F_2(x,Q^2)$ at
small $x$ was obtained in \cite{NZ_HERA} within the dipole
BFKL equation. The replacement of $\alpha_s(k_\perp^2)$ by
$\alpha_s(k_\perp^2/x(1-x))$ would obviously lead to a significant
change in the $x$ dependence of the proton structure
function. These facts indicate that $\alpha_s(k_\perp^2)$ rather than
$\alpha_s(k_\perp^2/x(1-x))$ should be used for the regime $z\gg L_f$.
It is noteworthy that the contribution of gluons with
$L_f\ll L_{QGP}$ at moderate energies is very significant. In
the oscillatory approximation $\sigma_{q\bar{q}}(\rho)\propto \rho^2$, it gives
the dominant contribution to the radiative parton energy loss 
 \cite{BDMPS1,LCPI2,Z_OA}.

The above consideration concerns the case of a low
density of the medium when the Landau-Pomeranchuk-Migdal effect is small. In the case of a dense
medium, the modified wavefunction $\Psi_{\{\lambda\}}^m(x,\ro,z)$ in
Eq. (\ref{eq:60}) should be calculated including any number of
rescatterings, which can be taken into account
by adding the imaginary potential $v=-in(z)\sigma_3(\rho,z)/2$
to the Hamiltonian (\ref{eq:50}). However, this does not change
the situation for the first decay vertex $q\to gq$, where
it is necessary to use $\alpha_s(k_\perp^2)$. In this case, in calculations 
in the coordinate representation with the use of
the diffusion relation (\ref{eq:20}), the typical  value of  
$k_\perp^2$ in $\alpha_s$ can
be related to the typical longitudinal distance dominating 
in the representation of $\Psi_{\{\lambda\}}^m(\ro,x,z)$
in terms of
the integral of the Green's function over z (details can
be found in \cite{Z04_RAA}). It is worth noting that thermal
effects in the dense QGP can result in the modification 
of $\alpha_s$ at virtualities of about several temperatures
of the QGP (this phenomenologically corresponds to
dependence of $\alpha_s$ on $z$). A parameterization of
$\alpha_s(Q^2)$ with freezing in the region of small $Q^2$ at
a certain value $\alpha_s^{fr}$ is widely used in current works on jet
quenching. For processes in vacuum, $\alpha_s^{fr}\approx 0.7-0.8$
\cite{DKT,NZ_HERA}. The analyses of the RHIC and LHC data on
the nuclear modification factor 
$R_{AA}$
\cite{RAA08,RAA11,RAA12,RAA13,Z_RPP} show
that the QGP can reduce 
$\alpha_s^{fr}$
to $\sim 0.4-0.6$. However,
this does not change the behavior of the running
charge in the region of large $k_\perp^2$, where the running charges  
for the two virtuality scales 
$k_\perp^2$ and $k_\perp^2/x(1-x)$ are still significantly different.

The quantitative difference in the induced gluon spectrum between 
two virtuality scales for $\alpha_s$ at the
decay vertex appears quite noticeable. In order to
quantitatively estimate the difference between these two
variants, we have calculated the gluon spectrum for the QGP with the
initial temperature $T = 400$ MeV at the proper time
$\tau_0 = 0.5$ fm (which approximately corresponds to central Pb + Pb 
collisions at $\sqrt{s}=2.76$ TeV at the LHC)
for the purely longitudinal expansion of the QGP within 
the $1 + 1$D Bjorken model. Our calculations within the 
LCPI approach (including all possible rescatterings) show that, in
the gluon energy range $\omega\sim 1-3$ GeV (dominating in
jet quenching) at an increase in the energy of the initial
parton from $20$ to $150$ GeV, the spectrum in the variants with 
$\alpha_s(k_\perp^2)$ and $\alpha_s(k_\perp^2/x(1-x))$
increases by $\sim 3-10$\%
and decreases by $\sim 23-30$\%, respectively. Because
of this difference in the behavior of the induced spectrum with  
increase of the initial parton energy, the nuclear modification 
factor $R_{AA}(p_T)$ for the
variant with $\alpha_s(k_\perp^2/x(1-x))$
grows with $p_T$ more
rapidly, which gives a better agreement with the LHC
data on $R_{AA}$ . However, our analysis shows that the variant 
with $\alpha_s(k_\perp^2/x(1-x))$ contradicts to well established
facts, and, hence, is theoretically unsatisfactory.

The method of incorporation of the running charge
by changing the fixed $\alpha_s$ 
in Eqs. (\ref{eq:100}), (\ref{eq:110}) to running $\alpha_s(k_\perp^2)$
is also not absolutely strict. The point is that the
possibility of separating the contribution associated with
evolution of $\alpha_s$ in higher radiative corrections to
Eqs. (\ref{eq:60}) and
(\ref{eq:90}) keeping the form of these formulas
is not obvious. Indeed, the inclusion of rescattering of
virtual $s$-channel gluons can result in the appearance
of interaction with
the medium  of the four-body parton states, which lead to a modification of the
evolution of $\alpha_s$ by the medium effects. Nevertheless, 
under the assumption that such effects are
small, and Eqs. (\ref{eq:60}), (\ref{eq:90}) provide a good approximation 
for the running charge as well, the above facts
indicate that $\alpha_s(k_\perp^2)$ is preferable.\\

This work was supported by the RFBR Grant No. 15-02-00668-a.\\

\end{document}